\documentclass[twocolumn,prl,showpacs,preprintnumbers,amsmath,amssymb]{revtex4-1}

\usepackage{graphicx}
\usepackage{dcolumn}
\usepackage{bm}

\begin{document}

\title{Comment on ``Multiorbital Effects on the Transport and the Superconducting Fluctuations in LiFeAs''}

\author{A. Ramos-\'Alvarez}
\author{J. Mosqueira}%
\author{F. Vidal}

\affiliation{LBTS, Departamento de Física da Materia Condensada, Universidade de Santiago de Compostela, E-15782 Santiago de Compostela, Spain}

\date{\today}


\pacs{74.70.Xa,74.25.F-,74.25.Jb,74.40.-n}

\maketitle     

In Ref.~\cite{Rullier} Rullier-Albenque \textit{et al.} measured the transverse magnetoresistivity $\delta\rho(H)/\rho(0)$ above the transition temperature $T_c$ in clean LiFeAs. These authors conclude that the conductivity induced by fluctuations, $\Delta\sigma$, follows a two-dimensional (2D) behavior even close to $T_c$, in spite that for LiFeAs the transverse coherence length $\xi_c(0)\approx1.6$~nm is larger than the Fe-layers spacing ($s=0.636$~nm), which would rather suggest a three-dimensional (3D) behavior. This proposal would have implications in the understanding of the multiband structure of iron pnictides, but it also contrasts with the 3D behavior observed near $T_c$ in the same compound \cite{Song} and in other iron pnictides with even smaller $\xi_c(0)/s$ \cite{Mosqueira}. Here we show that the proposal of Ref.~\cite{Rullier} could be an artifact associated to an inadequate subtraction of the normal-state (or \textit{background}) conductivity, $\sigma_B$.

Note first that in the clean crystals studied in Ref.~\cite{Rullier} $\sigma_B$ is orders of magnitude larger than the expected fluctuation contribution: at a reduced temperature $\varepsilon\equiv\ln(T/T_c)=10^{-1}$ the Aslamazov-Larkin (AL) approach predicts $\Delta\sigma_{\rm 2D}\sim2.5\times10^5\;\Omega^{-1}{\rm m}^{-1}$ and $\Delta\sigma_{\rm 3D}\sim1.5\times10^4\;\Omega^{-1}{\rm m}^{-1}$, whereas $\sigma_B\sim2\times10^7\;\Omega^{-1}{\rm m}^{-1}$ (note that in Ref.~\cite{Rullier} the AL $\Delta\sigma_{\rm 3D}$ is erroneously overestimated by a factor of 2). Thus, extracting $\Delta\sigma$ in these crystals would require a highly precise procedure to determine $\sigma_B$, which questions the adequacy of $\Delta\sigma$ to study the superconducting fluctuations in clean LiFeAs.

The procedure used in Ref.~\cite{Rullier} to determine the background conductivity assume a strict $H^2$ behavior of the magnetoresistivity in the normal state \cite{suple}. For temperatures near $T_c$, the deviation from this behavior observed at low fields is attributed to fluctuations. However, isotherms well above $T_c$, where fluctuation effects are negligible, present a similar $H^2$ dependence. This is difficult to appreciate in Fig.~2 of Ref.~\cite{Rullier} due to the scale, but may be clearly seen in the detailed view of the present Fig.~1 \cite{error}: isotherms above 45~K present a relative rounded behavior \textit{quite similar} to the one at 25~K, where fluctuation effects are claimed to be present. This shows that the $\delta\rho(H)/\rho(0)$ deviations from the $H^2$ behavior is a normal-state effect, that near $T_c$ will be superimposed to the superconducting fluctuation effects. 

Our analysis poses serious doubts about the conclusions drawn in Ref.~\cite{Rullier} about the 2D nature of fluctuation effects in LiFeAs. Moreover, it questions the applicability to this material of the model proposed for the quadratic dependence of $\delta\rho(H)/\rho(0)$ in the normal state.

This work was supported by the Xunta deGalicia (Grant No. GPC2014/038) and COST action MP1201 (NanoSC). A.R.-A. acknowledges financial support from Spain’s MICINN through FPI (Grant No. BES-2011-046820).

\begin{figure}[h]
\begin{center}
\includegraphics[width=.45\textwidth]{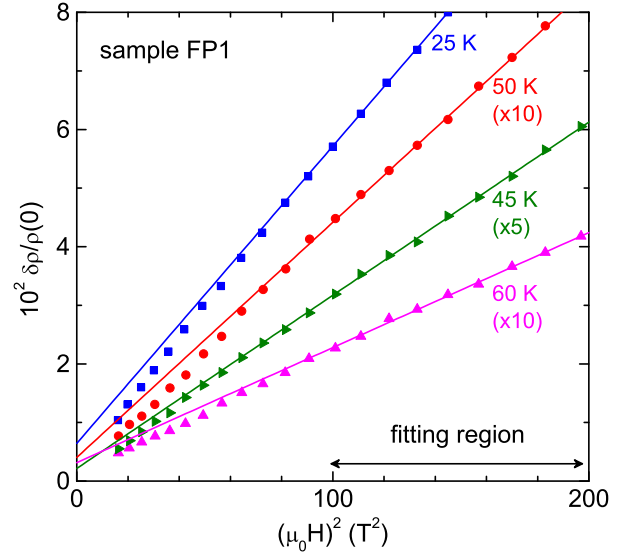}
\caption{Detail of the $H^2$ dependence of $\delta\rho/\rho(0)$ for sample FP1 at 25, 45, 50 and 60~K (for a better comparison some isotherms are multiplied by the indicated factor). For \textit{all} isotherms the lines are fits to the data above 100~T$^2$. These isotherms present a similar relative rounded behavior at low fields, in spite that fluctuation effects are assumed to be negligible above 45~K.}
\end{center}
\end{figure}


\vspace{-1cm}

\begin{references}

\bibitem{Rullier} F. Rullier-Albenque\textit{ et al}., Phys. Rev. Lett. {\bf 109}, 187005 (2012); \textit{ibid}.  {\bf 113}, 209901 (2014).

\bibitem{Song}Y.J. Song \textit{et al.}, Europhys. Lett. \textbf{97}, 47003 (2012).

\bibitem{Mosqueira}J. Mosqueira \textit{et al.}, Phys. Rev. B \textbf{83}, 094519 (2011).

\bibitem{suple}This is intended to be justified in the Supplementary Material for Ref.~\cite{Rullier}. Unfortunately, the claimed measurements of the Hall effect up to 14 T are not shown.

\bibitem{error}The uncertainty associated to the capture of the data points from Ref.~\cite{Rullier} through a standard graphical procedure remains below the data points size.


\end{references}
\end{document}